\documentclass[useAMS,usenatbib,twocolumn]{mn2e}

\usepackage{graphicx}
\usepackage{amssymb}

\title[Bound and unbound substructures in Galaxy-scale Dark Matter haloes]
      {Bound and unbound substructures in Galaxy-scale Dark Matter haloes}
\author[Maciejewski et al.]
{\parbox{18.0cm}{Michal Maciejewski$^{1}$\thanks{E-mail: michalm@mpa-garching.mpg.de}
, Mark Vogelsberger$^{1,2}$, Simon D.M. White$^{1}$,\\ Volker Springel$^{1,3}$}
\vspace{0.3cm}\\
$^{1}$Max-Planck-Institut f\"{u}r Astrophysik, Garching, Karl-Schwarzschild-Stra\ss e 1, 85741 Garching bei M\"{u}nchen, Germany\\
$^{2}$Harvard-Smithsonian Center for Astrophysics, 60 Garden Street, Cambridge, MA 02138, USA\\
$^{3}$Heidelberg Institute for Theoretical Studies, Schloss-Wolfsbrunnenweg 35, D-69118 Heidelberg, Germany}

\begin{document}

\pubyear{2010}

\pagerange{\pageref{firstpage}--\pageref{lastpage}} 

\maketitle
\label{firstpage}
\begin{abstract}

We analyse the coarse-grained phase-space structure of the six
Galaxy-scale dark matter haloes of the Aquarius Project using a
state-of-the-art 6D substructure finder. Within $r_{50}$, we find that
about $35\%$ of the mass is in identifiable substructures,
predominantly tidal streams, but including about $14\%$ in self-bound
subhaloes. The slope of the differential substructure mass function
is close to $-2$, which should be compared to $\sim -1.9$ for the
population of self-bound subhaloes. Near $r_{50}$ about $60\%$ of the
mass is in substructures, with about $30\%$ in self-bound subhaloes.
The inner $35$~kpc of the highest resolution simulation has only
$0.5\%$ of its mass in self-bound subhaloes, but $3.3\%$ in detected
substructure, again primarily tidal streams. The densest tidal
streams near the solar position have a 3-D mass density about $1\%$ of
the local mean, and populate the high velocity tail of the velocity
distribution.
\end{abstract}


\section{Introduction}

The detailed phase-space distribution of cold dark matter haloes can
substantially affect prospects for dark matter detection. Direct detection
experiments are starting to probe significant fractions of the parameter space
of plausible theoretical models, so a first detection of dark matter (DM) may
be imminent. Only recently has realistic prediction of the phase-space
structure near the Earth become possible, because very high-resolution
numerical simulations are required. It is now well established that the outer
regions of cold dark matter (CDM) haloes have a complicated phase-space
structure with many subhaloes and tidal streams \citep{Moore1999,Klypin1999,
 Ghigna2000, Stoehr2002,Diemand2004,Diemand2008,Springel2008}. This raises
the question of whether similar structures might affect direct detection
probabilities. Could the Earth be sitting in a ``hole'', i.e. a locally very
underdense region, in the DM distribution, as might occur in a fractal
structure, or if most of the mass near the Sun were concentrated in dense,
low-mass subhaloes. Some simulators have indeed argued that a significant
fraction of the local mass could lie in solar or Earth-mass subhaloes
\citep[e.g.][]{Diemand2005}, although more recent simulations suggest that the
local mass fraction in bound subhaloes of any mass is well below 1\%
\citep{Vogelsberger2009, Vogelsberger2010}. Other possibilities might be for
the Earth to lie within a subhalo, or within a dense tidal stream created by
disruption of an earlier subhalo. Either of these would produce a spike in
the velocity distribution of local dark matter particles.

Quantifying the detailed phase-space structure of a halo requires
disentangling its various DM components: the smooth component which is, in
fact, a superposition of many fundamental streams \citep{Vogelsberger2010};
compact, self-bound subhaloes; and tidal streams created by the disruption of
such subhaloes. Efficient identification of the tidal streams requires a
sensitive and robust structure-finder in 6D phase-space. Recently
\cite{Maciejewski2009a} presented an algorithm, Hierarchical Structure Finder
(HSF), designed specifically for this purpose. HSF identifies structures in an
N-body simulation as coherent, overdense sets of particles in the full 6D
position-velocity distribution. In this paper we use HSF to study the six
Galaxy-scale halos simulated as part of the Aquarius Project
\citep{Springel2008}, providing a quantitative analysis of the various halo
components and focussing, in particular, on the inner halo relevant for
detection experiments.

Our paper is organised as follows: Section \ref{sec:hsf} presents a brief
description of our Hierarchical Structure Finder. Section \ref{sec:str} then
explores the numerical convergence of our results by analysing simulations of
a single halo at a variety of resolutions. Finally we
use the full set of six Aquarius halos to study the expected scatter in
substructure properties among Galaxy-scale halos. Section \ref{sec:inn} 
focusses on substructure in the inner halo in order to assess possible
consequences for direct detection experiments. The final section gives our
conclusions.

\section{Hierarchical Structure Finder}
\label{sec:hsf}
Many different algorithms for identifying (sub)structure have been applied to
N-body simulations of the growth of cosmic structure. One of the first and
most widely used is the Friends-of-Friends (FOF) scheme introduced by
\cite{Davis1985}; this defines ``halos'' as disjoint particle sets containing
every particle closer than some maximal linking length to at least one other
member of its set. This bounds objects approximately by an isodensity
surface, but makes no assumption about their shape or internal structure. In
contrast, the Spherical Overdensity (SO) group-finder of \cite{Cole1996} finds
high-density peaks in the particle distribution, grows spheres centred on each
until the mean enclosed density drops to a specified value (typically $\sim
200$ times the cosmological mean) and then defines halos as the contents of
those spheres whose centres do not lie within a more massive halo. More
recent structure-finders [e.g. SKID \citep{Governato1997}, SUBFIND
 \citep{Springel2001}, VOBOZ \citep{Neyrinck2005}, PSB \citep{Kim2006},
 ADAPTAHOP \citep{Aubert2004}, AHF \citep{Knollmann2009}, 
HSF \citep{Maciejewski2009a}] typically identify objects as connected self-bound
particle sets above some density threshold. Such methods have two steps. 
Each particle is first linked to a local DM density maximum by following 
the gradient of a particle-based estimate of the underlying DM density field. 
The particle set attached to a given maximum defines a candidate structure. 
In a second step, particles which are gravitationally unbound to the structure
are discarded until a fully self-bound final object is obtained. The various 
methods differ in the way particles are treated when they belong to more than 
one candidate and in the way unbound particles are redistributed. Most methods 
produce a hierarchical characterisation of structure where halos contain subhaloes
which in turn can contain their own subhaloes.

These methods can be extended to higher dimensions, in particular to
6D phase-space. The main complication is then that the smoothed
density field and its gradient must be estimated from the particle
distribution in six dimensions. The Hierarchichal Structure Finder
(HSF) presented by \cite{Maciejewski2009a} is an algorithm of this
type, and can be used, just like the above 3D algorithms, to identify
bound subhaloes. Other six dimensional phase-space structure finders
have been developed recently by \citet[][6dFOF]{Diemand2006} and
\citet[][EnLink]{Sharma2009}. In the following we describe the HSF method
in somewhat more detail, since this is the method we will use for the
rest of this paper. Further technical details and tests can be found
in \cite{Maciejewski2009a}.

To find candidate structures we first need to estimate phase-space densities
at the positions of all the particles. Furthermore we need to calculate local
phase-space density gradients. HSF does this using a six-dimensional SPH
smoothing kernel with a local adaptive metric as implemented in the EnBiD code
\citep{Sharma2006}. Neighbouring particles can then be used to derive the
required gradients. For the SPH kernel we use $N_{\rm sph}=64$ neighbours
whereas for the gradient estimate we use $N_{\rm ngb}=20$ neighbours.

Once phase-space densities have been calculated, we sort the particles
according to their density in descending order. Then we start to grow
structures from high to low phase-space densities. While walking down in
density we mark for each particle the two closest (according to the local
phase-space metric) neighbours with higher phase-space density, if such
particles exist. In this way we grow disjoint structures until we encounter a
saddle point, which can be identified by observing the two marked particles
and seeing if they belong to different structures. A saddle point occurs at
the border of two structures. In the standard setup of HSF, which is used
throughout this paper, the masses of the two structures separated by the
saddle point are compared and the smaller one is cut, defining a complete
individual structure. All particles below the saddle point whose higher
density neighbours are part of the cut object are attached to the other,
larger structure. Pursuing this procedure until all particles
have been considered divides a halo into a unique disjoint set of
substructures, of which the most massive, which also contains the lowest
phase-density particles, is the main substructure.

When we wish to isolate self-bound subhaloes, we follow an identical
procedure, except that each time we reach a saddle point, we remove
all unbound particles iteratively from the smaller structure and
attach them provisionally to the larger structure. Once we have
followed this algorithm down to the lowest phase-density particle, we
are left with a set of self-bound subhaloes and a few particles which
are bound to no subhalo, not even the most massive self-bound subhalo
which again is the one containing the lowest phase-density bound
particles.

Both these procedures divide a halo into a disjoint set of phase-space
structures, each containing a single phase-space density peak and
bounded approximately by a level surface of phase-space density. In
the first procedure each structure normally contains both bound and
unbound particles, and all halo particles are assigned to some
structure. In the second procedure, each structure is self-bound, and
some halo particles are not assigned to any structure. To be
specific, in the following we will refer to all particles inside
$r_{50}$\footnote{This is defined as the radius of the largest sphere
centred on the halo density peak which encloses a mean density at
least $50$ times the critical value.} as the {\em halo}. We call the
most massive substructure constructed from these particles the {\em
main halo}. Note that by definition it cannot extend beyond $r_{50}$
and that its mean density within $r_{50}$ will be less than 50 times
the critical density. Note also that the main halo will change
slightly according to whether we do or do not apply the unbinding and
reassignment procedures. In the former case we refer to all other
structures as (self-bound) {\em subhaloes}, whereas in the latter case
we refer to them as {\em substructures}. A subhalo is thus always
part of a substructure, but a substructure does not necessarily
contain a subhalo.

\section{Structures in the Aquarius simulations}
\label{sec:str}

We study the phase-space structure of Milky Way-sized DM haloes using
the high-resolution simulations of the Aquarius Project
\citep{Springel2008}. The cosmological parameters for these
simulations are $\Omega_m=0.25, \Omega_\Lambda = 0.75 ,\sigma_8=0.9$
and $H_0=73 \rm{km}$ ${\rm s}^{-1}\rm{Mpc}^{-1}$. For this project
six Galaxy-mass haloes (Aq-A to Aq-F) were selected from a lower
resolution version of the Millennium-II Simulation \citep{Boylan2009}
and resimulated with progressively higher particle number and smaller
softening length. The haloes were selected to have no close massive
companion at $z=0$. When studying differences in phase-space structure
between these haloes, we use the second resolution level (the highest
for which results are available for all six objects). At this
resolution all haloes have more than $1.6\times10^8$ particles inside
$r_{50}$, corresponding to a particle mass $\sim10^4{\rm M}_{\sun}$.
In addition, we use resimulations of the Aq-A halo at four different
resolution levels to check the numerical convergence of our results.
In the final section of this paper we investigate phase-space
structure in the inner halo, defined as $r<r_{\rm inner}=35$~kpc. For this
purpose we use three resolution levels of the Aq-A halo with the
largest one (Aq-A-1) having almost $1.5\times10^9$ particles inside
$r_{50}$ and more than $2\times10^8$ particles inside $r_{\rm inner}$.
Together with the ability of HSF to analyse the full 6D particle
distribution, this simulation set allows the first robust and fully
general quantification of the various phase-space components predicted
by the $\Lambda$CDM model at $r\sim 8$~kpc where direct detection
takes place.

\subsection{A resolution study}
\label{sec:res}

\begin{figure}
\includegraphics[width=8.5cm,height=8.5cm]{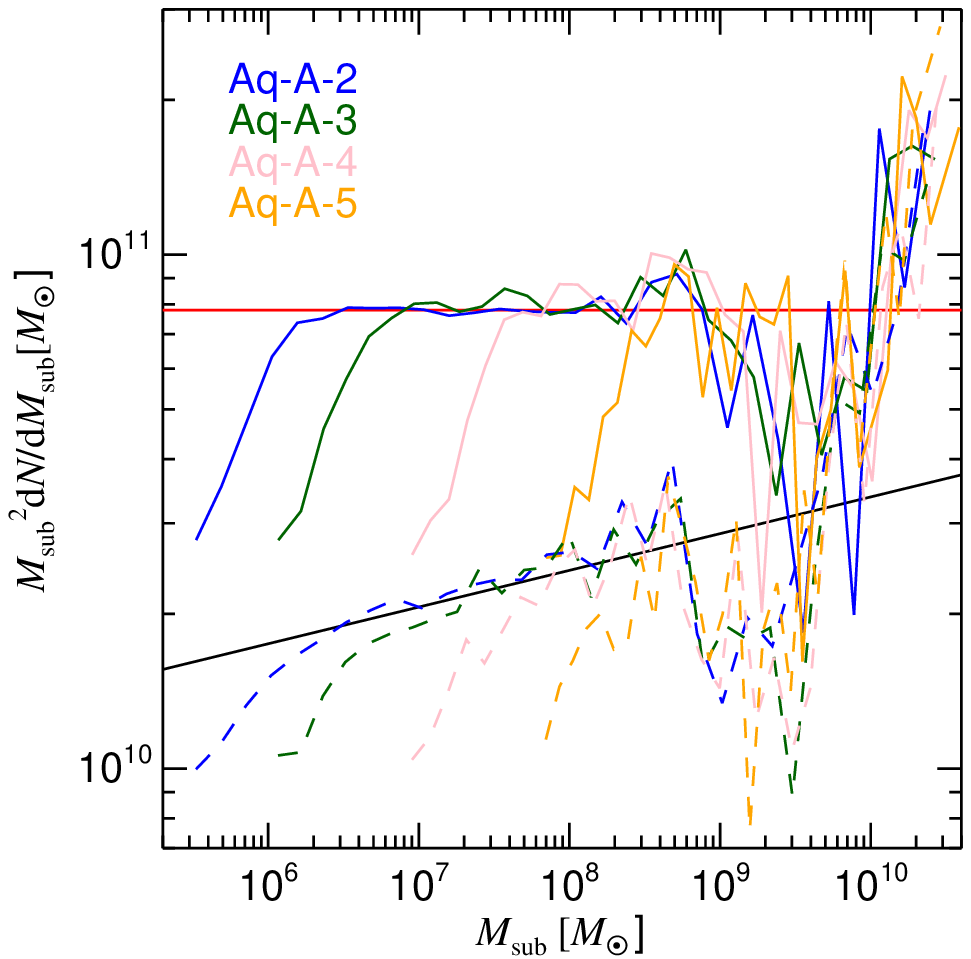}
\includegraphics[width=8.5cm,height=8.5cm]{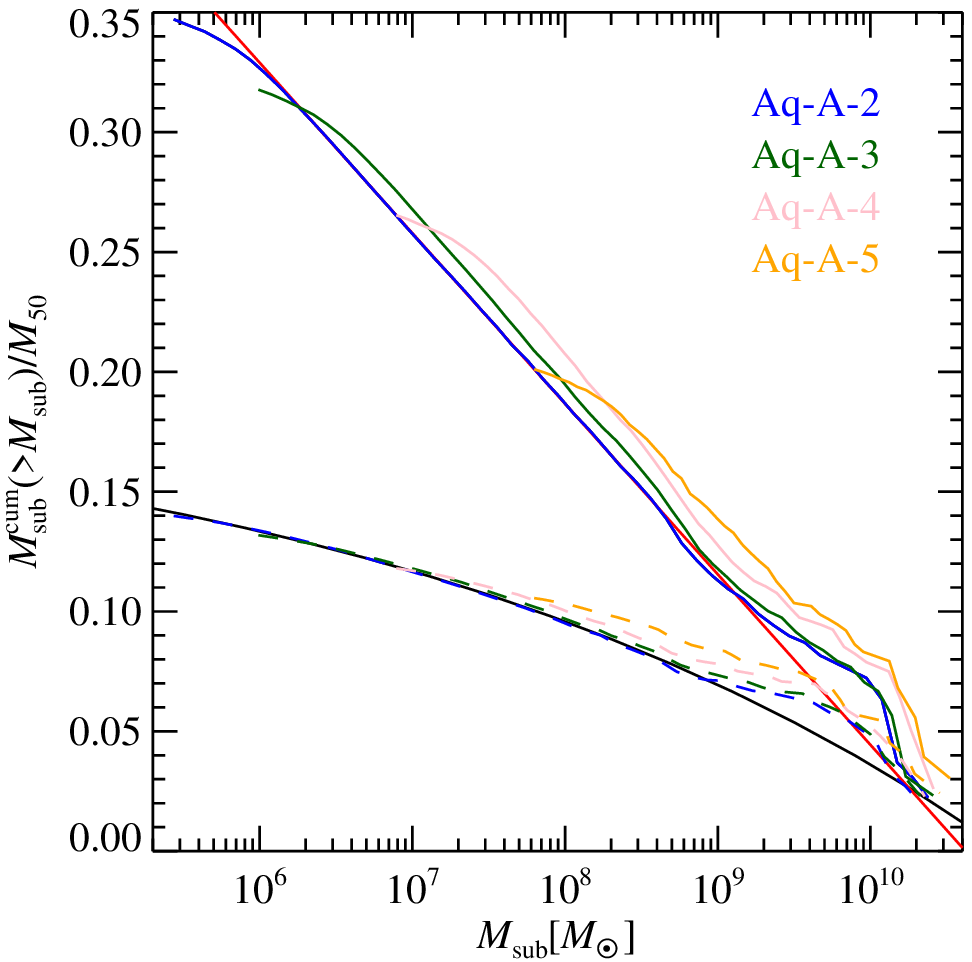}
\caption{Top panel: Differential substructure mass functions for
  different resolutions of the Aq-A halo in the region
  $r<r_{50}=433$~kpc. Solid lines indicate mass functions for the
  substructures identified when HSF is executed without unbinding
  and reassignment procedures (see the text), while dashed lines
  indicate the corresponding functions for the self-bound subhaloes
  found when these procedures are implemented. Different colours refer
  to the different resolution levels as indicated in the plot. The
  straight black line is a power-law fit, $\mbox{d}N/\mbox{d}M\propto
  M^{-1.9}$, to the data for self-bound subhaloes. The data for
  substructures are instead fit by a power-law of slope $-2$ (the
  horizontal red line). Bottom panel: Corresponding cumulative mass
  functions inside $r_{50}$, expressed as the total fraction of the
  enclosed mass in identified substructure. Line styles are the same
  as in the top panel. Black and red lines are based on the fits in
  the upper panel with a high mass truncation at $2\%$ of the total
  mass.}
\label{mass_res}
\end{figure}

We begin by analysing the mass functions of substructures and of
self-bound subhaloes in the Aq-A halo and their dependence on
resolution. To be consistent with earlier work we define the edge of
the halo at $r_{50}=433$~kpc and we count all objects within this
radius, but we note that this is a large radius and, as a result, the
counts are dominated by objects beyond 100~kpc, more than an order of
magnitude further from the Galactic Centre than the Sun. In the upper
panel of Fig.~\ref{mass_res} we compare the differential mass
functions of substructures (solid curves) and of self-bound subhaloes
(dashed curves) at four different resolutions. The lower panel shows
the corresponding cumulative mass functions. In both cases the mass
functions agree quite well between the simulations above their
respective resolution limits. For the self-bound subhaloes, the slope
of the differential mass function is close to $-1.9$, as found earlier
in the SUBFIND analysis of \cite{Springel2008}. At lower resolution
the distribution is better approximated, particularly in the low mass
bins, by a slope close to $-1.8$. For the Aq-A-2 halo, we find that
$14\%$ of the halo mass is in self-bound subhaloes, which is 20\%
higher than the corresponding SUBFIND value \citep[$12\%$
  -][]{Springel2008}, reflecting the fact that HSF typically attaches
more mass to each identified object than SUBFIND. In the lower
resolution simulations, particles from unresolved low-mass
substructures are in many cases attached to more massive objects.
This explains the shifts in the mass functions at high masses that are
well resolved by all simulations (see the bottom panel of
Fig.\ref{mass_res}). The mass fraction in self-bound subhaloes
changes from $11\%$ for Aq-A-5 to $14\%$ for Aq-A-2.

The power-law behaviour of the mass function of self-bound subhaloes has been
known for some time, but there has been controversy over its slope. If this
slope is $-2$, then the mass fraction in subhaloes diverges logarithmically at
low mass, and is cut off at a mass corresponding to the free-streaming length of
the underlying DM particle (typically Earth mass for neutralino candidates).
If the slope is -1.9, however, as found above, then the mass fraction in
subhaloes has already effectively converged at the limit of the highest
resolution Aquarius simulations and hence is $\sim15\%$ within
$r_{50}$. Within smaller radii this fraction drops dramatically, as we will
see below.

HSF makes it possible to find unbound substructures also, and
the solid lines in Fig. \ref{mass_res} show that within $r_{50}$ rough
numerical convergence is achieved for their mass function. In this case,
however, the slope appears close to $-2$ and the mass in substructures exceeds
that in self-bound subhaloes by more than a factor of 2 at all masses, and by
increasingly large amounts at small mass. To the resolution limit of Aq-A-2,
35\% of the mass within $r_{50}$ is in substructure, showing the total mass
detected in unbound tidal streams to be significantly larger than in
self-bound subhaloes. With increasing resolution, significantly more
substructures are found, and mass shifts from massive to smaller
substructures, as found above for self-bound subhaloes but even more
strongly. The behaviour seen in the lower panel of Fig.~\ref{mass_res}
cannot be extrapolated straightforwardly to lower mass. As we will see below,
at the resolution of Aq-A-2, most of the mass in the outer halo is resolved
into substructures, but relatively little of the mass in the inner regions. At
higher resolution it will be the transition between these two regimes which
controls the total mass fraction in substructure, rather than the increase in
resolved substructures at any particular radius.

\begin{figure}
\includegraphics[width=8.5cm,height=8.5cm]{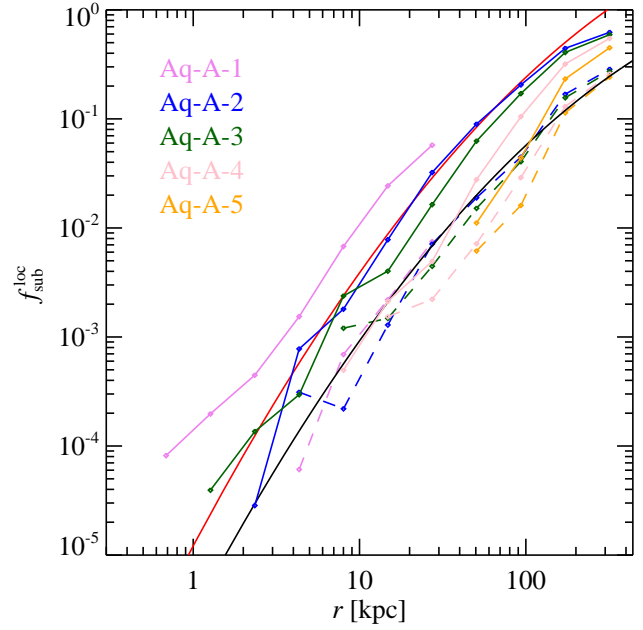}
\caption{Fraction of mass in substructures and in self-bound subhaloes
  as a function of radius estimated from a set of disjoint spherical
  shells and for various resolution levels of the Aq-A halo as
  indicated by colour. We take the mass within $r_{50}$ for the 4 lower
  resolution haloes. For Aq-A-1 only the mass within the inner $35$~kpc was used.
  Solid lines represent the substructures and
  dashed lines the self-bound subhaloes. The black line is an analytic
  fit based on data for the full set of 6 Aquarius haloes (see below)
  $f_{\rm sub}^{\rm loc}=\exp[\gamma+\beta
    \ln(r/r_{50})+0.5\alpha\ln^2(r/r_{50})]$ with parameters
  $\alpha=-0.31,\beta=0.98$ and $\gamma=-1.09$. The red line is the
  same function shifted vertically to $\gamma=0.35$.}
\label{radius_res}
\end{figure}

For this same resolution series of Aq-A simulations,
Fig.~\ref{radius_res} shows the mass fractions in substructures and in
self-bound subhaloes for a set of of 10 disjoint spherical shells
extending from 1~kpc to $r_{50}$. Within 35 kpc we also show results for
the highest resolution simulation Aq-A-1. Although the results for a single
simulation are rather noisy, they can be represented reasonably well
by
\begin{equation}
f_{\rm sub}^{\rm loc}=\exp[\gamma+\beta \ln(r/r_{50})+0.5\alpha\ln^2(r/r_{50})],
\label{eq:fsub}
\end{equation}
with parameters $\alpha=-0.31,\beta=0.98$ and $\gamma=-1.09$ for the
subhaloes. This analytic form was used to fit the radial
distribution of SUBFIND subhaloes in \cite{Springel2008}. We note
that the quoted parameters were obtained from a fit to data for the
full set of six level 2 Aquarius haloes (see below). The distribution
of HSF subhaloes is very similar to that of SUBFIND subhaloes, but HSF
attaches slightly more particles to objects near the centre.

We find substructures down to 0.6~kpc in Aq-A-1 and down to $2$~kpc in Aq-A-2
and Aq-A-3 simulations. For Aq-A-4, however, no substructures are found
within $\sim 9$~kpc, demonstrating that at least $\sim10^7$ particles are
needed within $r_{50}$ to begin to study streams around the Sun's position.
The red curve shows the prediction of Eq.~(\ref{eq:fsub}) for parameters
 $\alpha=-0.31,\beta=0.98$ and $\gamma=0.35$ but, in contrast to the
situation with subhaloes, it is clear that the results in the inner regions
are not converging with increasing resolution. The results for Aq-A-1 lie
well above the red line and should clearly still be considered as a lower
limit to the mass fraction contained in tidal streams in these regions.

\subsection{Mass distribution inside $r_{50}$}
\label{sec:r50}

\begin{table*}
\begin{tabular}{|c|c|c|c|c|c|}\hline
Halo & SUBFIND & $N_{\rm SUBFIND}$ & HSF subhaloes & $N_{\rm HSF}$ & HSF substructures \\ 
     & (per cent)&                 & (per cent)    &               & (per cent) \\ \hline
Aq-A-2 & $12.16$ & $45024$ & $14.14$ & $ 48052$ & $34.71$\\
Aq-B-2 & $10.54$ & $42537$ & $14.44$ & $ 44143$ & $33.65$\\
Aq-C-2 & $7.17$  & $35022$ & $7.72$  & $ 36525$ & $29.60$\\
Aq-D-2 & $13.06$ & $47014$ & $14.26$ & $ 49726$ & $34.49$\\
Aq-E-2 & $10.75$ & $42725$ & $13.10$ & $ 44400$ & $33.65$\\
Aq-F-2 & $13.39$ & $52503$ & $15.16$ & $ 57269$ & $34.18$\\ \hline
\end{tabular}
\caption{Total mass in substructures within $r_{50}$ for the 6 Aquarius haloes
  and different structure-finding methods. Mass fractions are calculated
  relative to the total mass within $r_{50}$. {\em Halo:} Aquarius halo
  label, {\em SUBFIND:} total mass fraction of particles in bound subhaloes
  found by SUBFIND; {\em $N_{\rm SUBFIND}$:} number of bound subhaloes found
  by SUBFIND; {\em HSF subhaloes:} total mass fraction of particles in bound
  subhaloes found by HSF; {\em $N_{\rm HSF}$:} number of bound subhaloes found
  by HSF; {\em HSF substructures:} total mass fraction of particles in substructures found
  by HSF.}

\label{table2}
\end{table*}

\begin{figure}
\includegraphics[width=8.5cm,height=8.5cm]{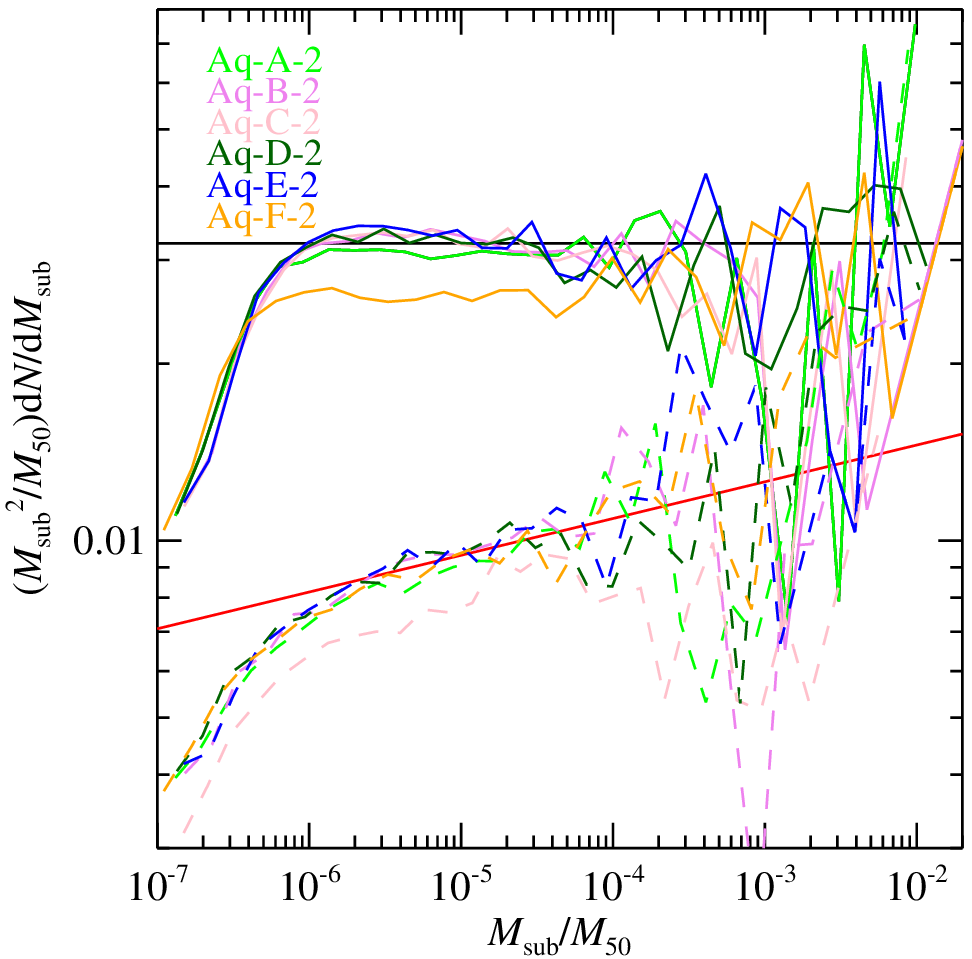}
\includegraphics[width=8.5cm,height=8.5cm]{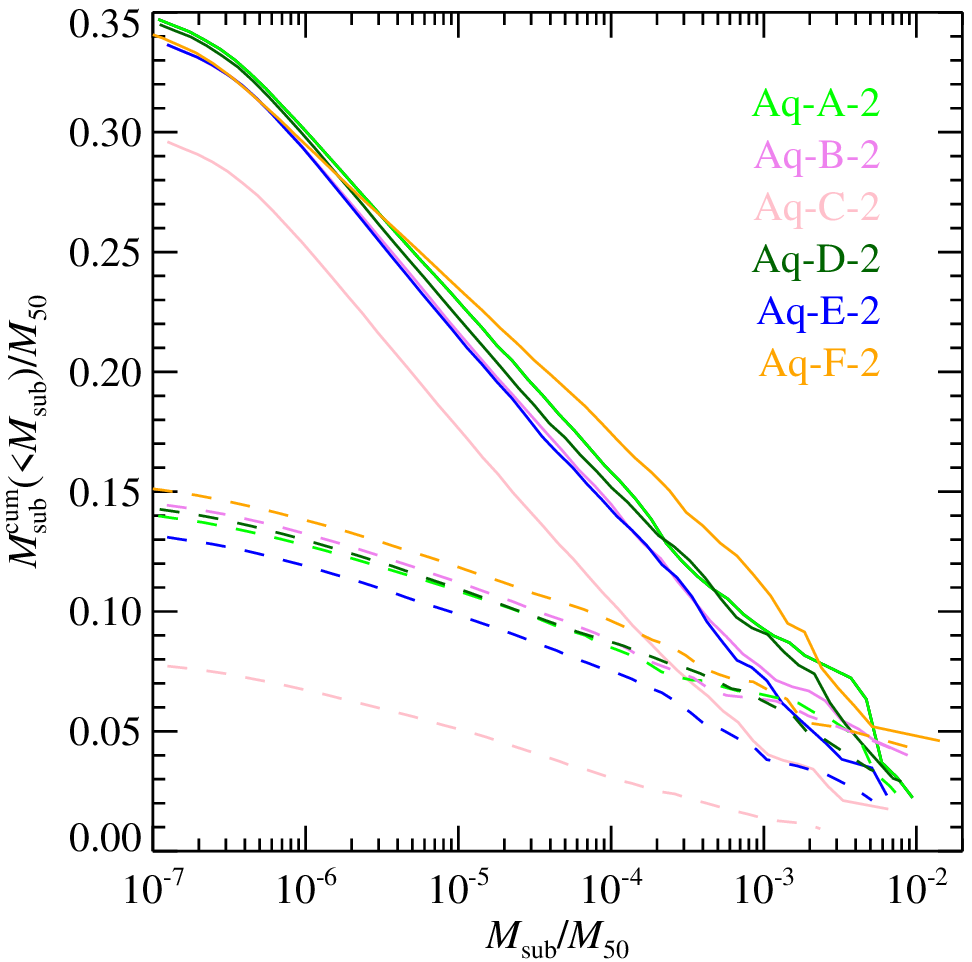}
\caption{Top panel: Differential subhalo mass function for the six different
  Aquarius haloes Aq-A-2 to Aq-F-2. Solid lines show substructures and dashed
  lines self-bound subhaloes. The red line shows a power law
  $\mbox{d}N/\mbox{d}M\propto M^{-1.9}$ which is a good fit to the
  differential mass function of the self-bound subhaloes. The black line shows
  a $-2$ power-law which better describes the differential mass function of
  substructures. Bottom panel: The corresponding cumulative mass functions.
  Line styles are the same as in the top panel.}
\label{mass_profile}
\end{figure}

\begin{figure}
\includegraphics[width=8.5cm,height=8.5cm]{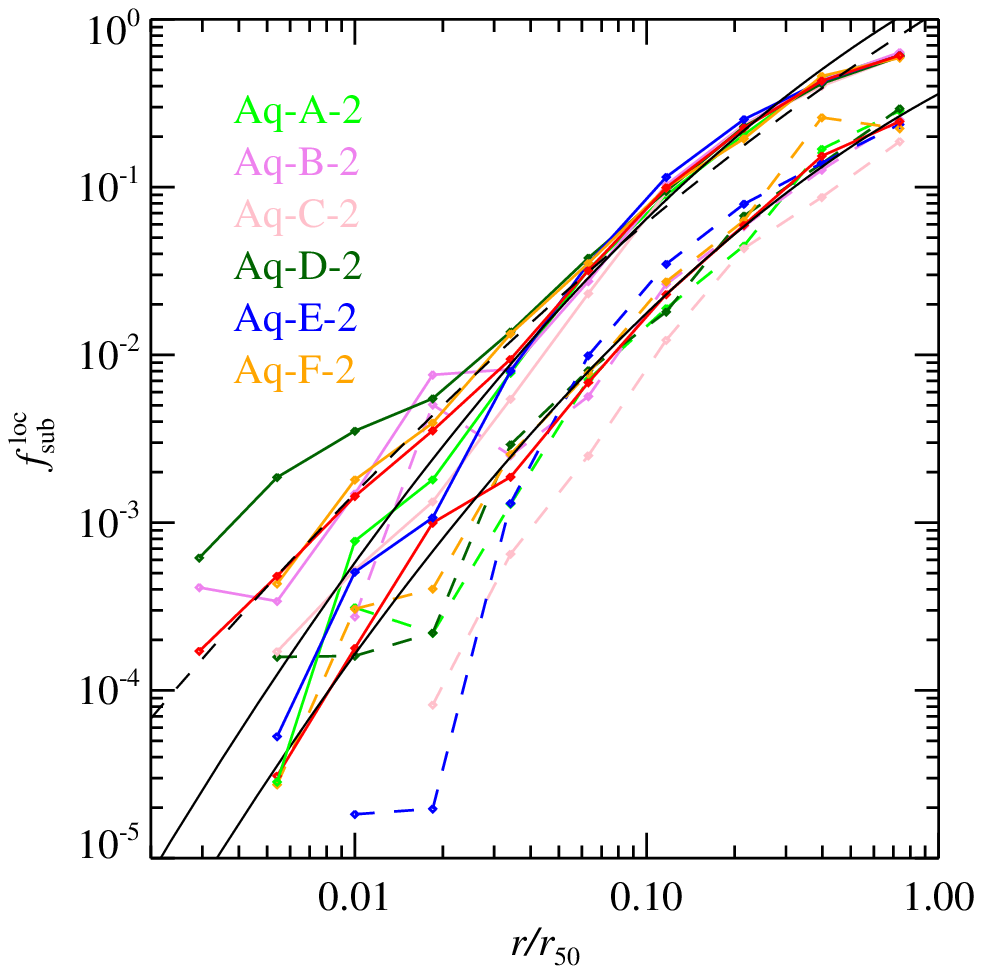}
\caption{Fraction of mass in substructure as a function of radius for the six
  different haloes Aq-A-2 to Aq-F-2. Solid lines are for all substructures and
  dashed lines are for self-bound subhaloes. Red lines show the mean of the
  halo sample and black lines show analytic fits using the function
  $f_{\rm sub}^{\rm loc}=\exp[\gamma+\beta
    \ln(r/r_{50})+0.5\alpha\ln^2(r/r_{50})]$ with parameters
  $\alpha=-0.31,\beta=0.98$ and $\gamma=-1.09$ for subhaloes and the same
  function shifted vertically to $\gamma=0.35$ for substructures. The black
  dashed line shows the best fit for substructures with parameters
  $\alpha=-0.16,\beta=1.10$ and $\gamma=0.10$.}
\label{mass_radius}
\end{figure}

In this section we quantify the object-to-object scatter in the substructure
mass function by analysing all six level 2 haloes of the Aquarius Project. Two
of our haloes Aq-A and Aq-C did not experience major mergers below redshift
$3$. Haloes Aq-B and Aq-F on the other hand each underwent a major merger
below redshift $1.5$. More information on the merger history of the Aquarius
haloes and how representative these haloes are of the population of Milky
Way-like haloes can be found in \cite{Boylan2009} and in \cite{Wang2010}.

The top panel of Fig.~\ref{mass_profile} shows the differential substructure
mass function for the six Aquarius haloes Aq-A-2 to Aq-F-2; the bottom panel
presents corresponding cumulative mass functions. Table~\ref{table2} lists
values for the total mass fraction in substructure in each halo using
different substructure identification methods. SUBFIND subhaloes
\citep{Springel2008} and subhaloes found by HSF follow the same power-law with
a slope close to $-1.9$. Halo Aq-C-2 evolves in a quiet merger environment and
this explains its deficit in substructures. For the general substructures the
slope of the power-law is close to $-2$, but it is difficult to measure this
value accurately because the low-mass end is contaminated by substructure
arising through discreteness noise, particles which are connected by HSF but
do not represent physical substructures.

In Aq-C-2 only $8\%$ of the mass within $r_{50}$ belongs to self-bound
subhaloes, while for Aq-E-2 this number is $13\%$ and for the other haloes it
is $14-16\%$. The subhalo mass of Aq-F-2 is dominated by the largest
subhaloes. It is interesting to observe that although Aq-C-2 has the fewest
self-bound subhaloes, its mass fraction in substructures is about $30\%$,
close to the value for Aq-B-2 ($34\%$) which had completely different and much
richer merger history. All the other haloes also have substructure mass
fractions around $34\%$. The slope of the differential substructure mass
function is similar in all haloes except Aq-F-2, where the very recent merger
apparently causes a bias towards massive substructures.

Fig.~\ref{mass_radius} shows the fractions of mass in substructures (solid
lines) and in self-bound subhaloes (dashed lines) as a function of distance
from halo centre. For most of the haloes the latter follows Eq.~(\ref{eq:fsub}),
indicated as a black solid line. Clearly, HSF identifies substructures close
to the centre in all halos. This is possible because of the high density
contrast in 6D phase-space compared to 3D configuration space.
Eq.~(\ref{eq:fsub}) with a different normalisation (i.e. a vertical shift, see
the solid black line) also gives a rough fit to the radial dependence of the
substructure mass fraction in most haloes, but we note that an independent fit
of the same functional form (the black dashed line) suggests that the mass
fraction in substructures increases relative to that in self-bound subhaloes
in the inner regions of the haloes.

In the outskirts of the haloes, near $r_{50}$, up to $30\%$ of the mass
is in the form of self-bound subhaloes and up to $60\%$ resides in
substructures.

\section{The inner halo}
\label{sec:inn}

For many years experimenters have been trying to detect DM in laboratory
devices. Detector signals are very sensitive to the local DM phase-space
distribution, so it is important to study halo structure in detail at the
solar position. This requires high-resolution simulations like those of the
Aquarius Project. A first phase-space analysis using the Aquarius haloes was
carried out by \cite{Vogelsberger2009} with a focus on the local velocity and
spatial distributions and their imprints on direct detection signals. Here we
extend this study and focus on phase-space structures at $r\sim 8$~kpc. We go
beyond the self-bound subhalo analysis of \cite{Springel2008} by using HSF,
which efficiently identifies gravitationally unbound structures like tidal
streams. Such features can have a significant impact on DM experiments, and
our goal here is to quantify the total amount of structure in the inner halo,
both bound and unbound. We therefore concentrate on the region within $r_{\rm
  inner}=35$~kpc and use the excellent resolution of Aq-A-1 to analyse
structures near the solar circle.

\subsection{Substructure in the inner halo}
\label{sec:subinn}

\begin{figure}
\includegraphics[width=8.75cm,height=10.5cm]{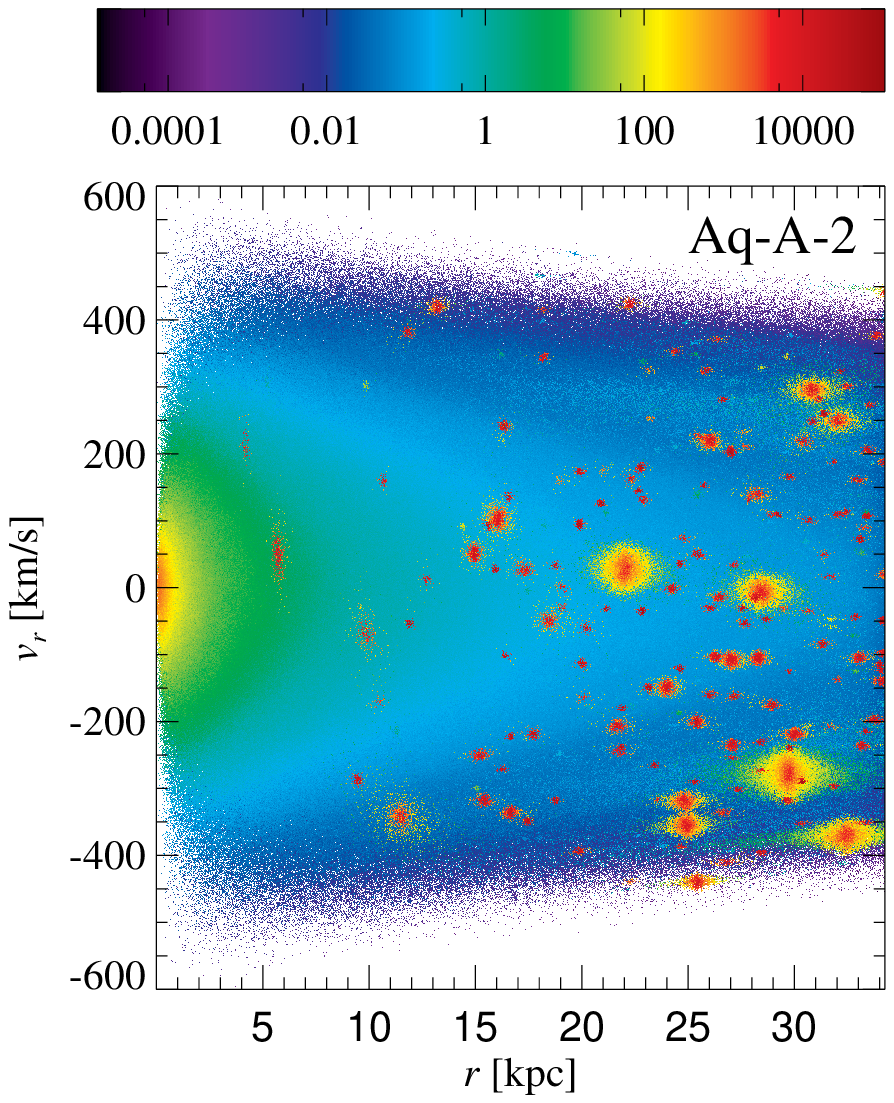}
\includegraphics[width=8.75cm,height=10.5cm]{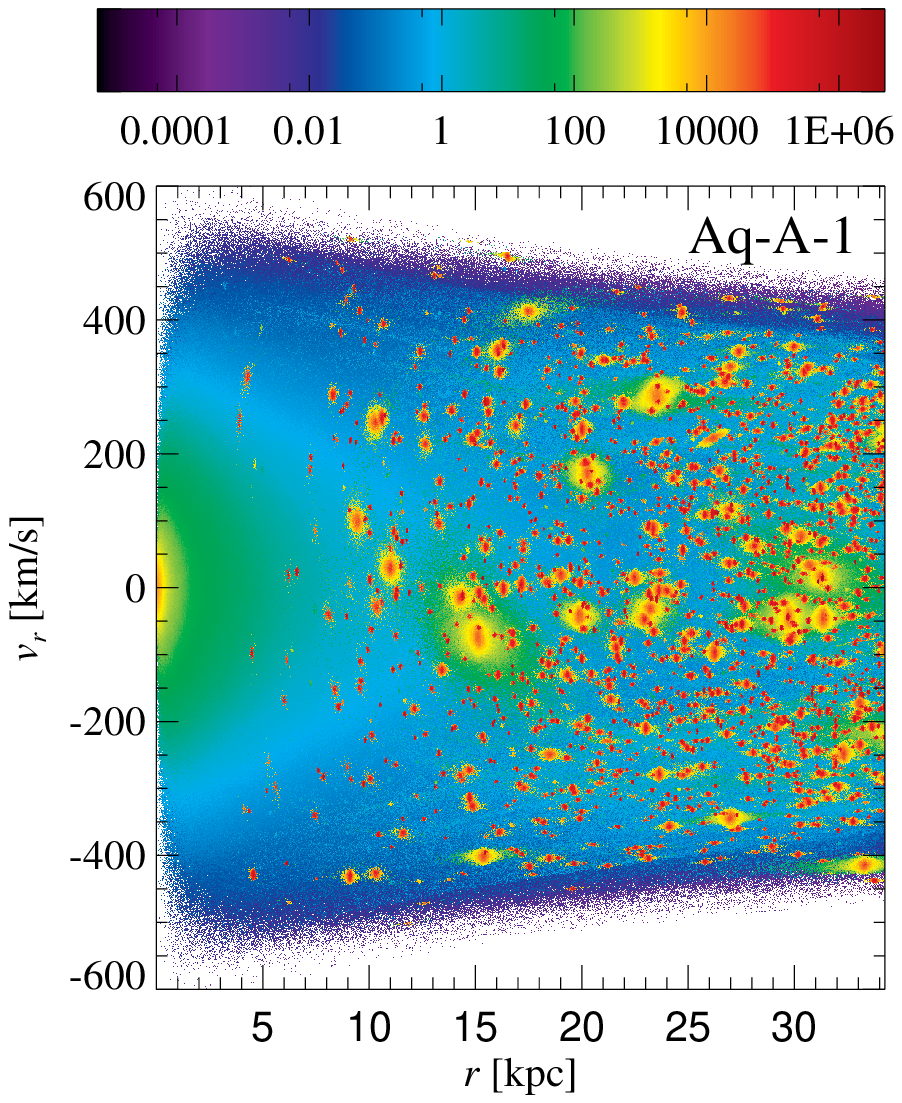}
\caption{Phase-space $r$-$v_r$ plots of Aq-A-2 (top) and Aq-A-1 
  (bottom). These were constructed by first calculating the
  phase-space density at each particle using EnBiD \citep{Sharma2006}. We then create 
  a pixelised image with
  $1000\times1000$ bins and for each pixel we calculate the logarithm of the
  maximum phase-space density over all particles in that pixel. The
  phase-space density is measured in units of
  $M_{\sun}~\mbox{kpc}^{-3}\mbox{km}^{-3}~\mbox{s}^3$.}
\label{rvr}
\end{figure}

To give a first impression of phase-space structure in the inner halo we use
the technique of \cite{Maciejewski2009b}. We estimate the phase-space density
at the position of each particle with EnBiD \citep{Sharma2006} and plot an $r$-$v_r$
phase-space portrait in which each pixel is colour-coded according to the
maximum phase-space density of the particles it contains. The top panel of
Fig.~\ref{rvr} shows the resulting phase-space plot for the inner part of the
Aq-A-2 halo, while the bottom panel shows the corresponding plot for Aq-A-1.
Aq-A-2 has about $2.5\times10^7$ and Aq-A-1 about $2\times10^8$ particles
inside $r_{\rm inner}$. The increased resolution results in substantially
more self-bound structures and tidal streams being visible in the inner
regions of Aq-A-1. In the following we quantify these phase-space structures
in some detail.

\begin{figure}
\includegraphics[width=8.5cm,height=8.5cm]{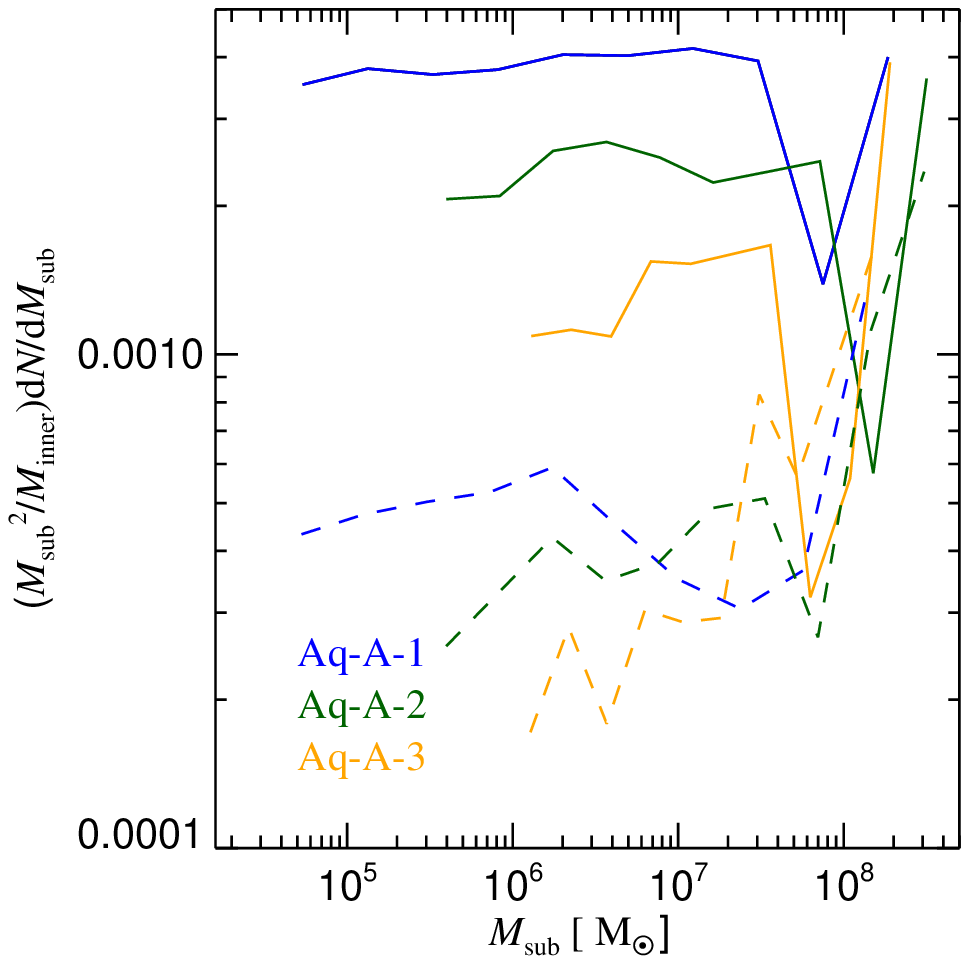}
\caption{Differential mass functions for substructures and for self-bound
  subhaloes in the inner part of the Aq-A halo ($r < r_{\rm inner}=35$~kpc) at
  various resolutions. Solid lines show results for substructures and dashed
  lines for self-bound subhaloes.}
\label{mass_profile_inner}
\end{figure}

\begin{figure}
\includegraphics[width=8.5cm,height=8.5cm]{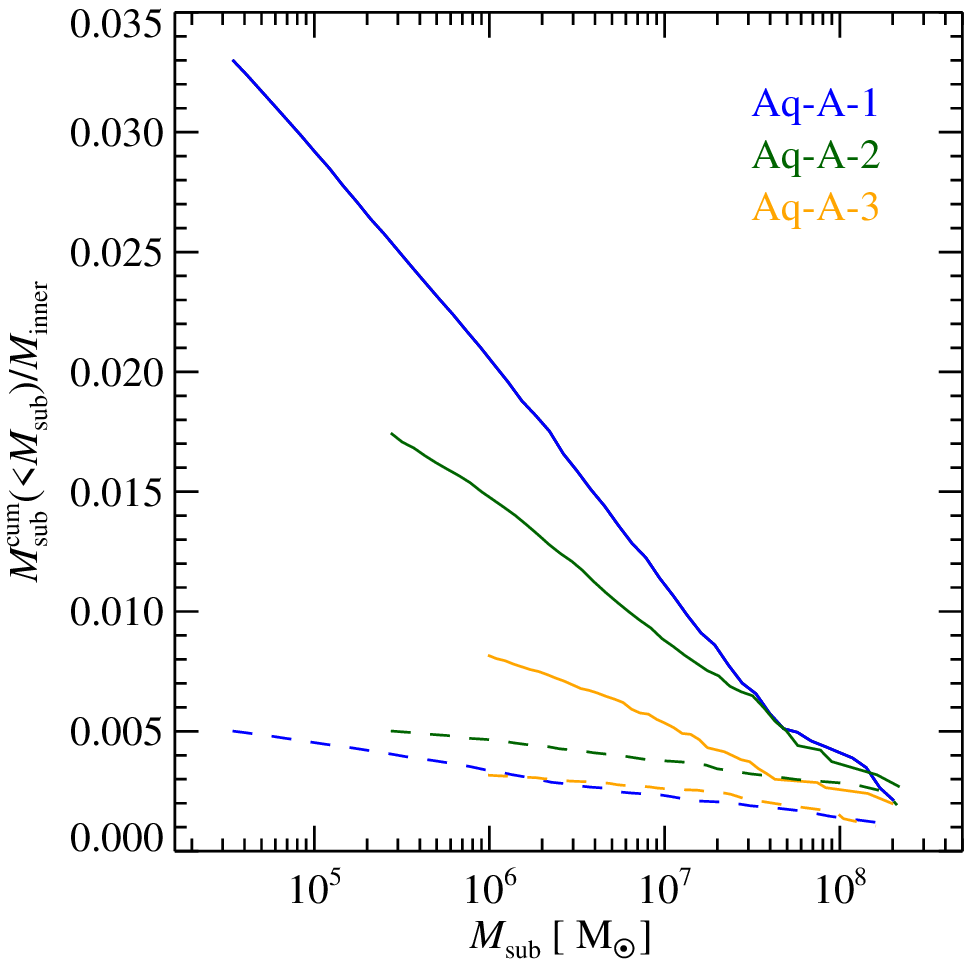}
\caption{Cumulative mass functions for the inner part of the Aq-A halo ($r <
  r_{\rm inner}=35$~kpc) at different resolutions. Solid lines represent
  substructures and dashed lines self-bound subhaloes. All curves are
  normalised to $M_{\rm inner}$, the mass within $r_{\rm inner}$.}
\label{mass_cum_inner}
\end{figure}

The solid lines in Fig.~\ref{mass_profile_inner} show the differential mass
functions of the substructures found by HSF within $r_{\rm inner}$ for the
three highest resolution Aq-A haloes. The number of substructures more
massive than $10^7M_\odot$ is quite small and only for Aq-A-1 is the dynamic
range sufficient to determine a power-law slope, which is close to $-2$.
While the differential mass function for substructures within $r_{50}$ (the
top panel of Fig.~\ref{mass_profile}) converges reasonably well with
increasing resolution, this is not the case for the inner regions. Here,
increasing resolution enables tidal streams to be followed to significantly
lower contrast, substantially increasing the mass attached to each one and so
the total mass in substructures. This effect was already visible in
Fig.~\ref{radius_res} and is confirmed by the cumulative mass function shown
in Fig.~\ref{mass_cum_inner}. Here also the curves for different resolutions
agree much less well than was the case when we focused on substructures within
$r_{50}$ (see the bottom panel of Fig.~\ref{mass_profile}).

Only $0.5\%$ of the mass inside $r_{\rm inner}$ is in the form of self-bound
subhaloes. Although this number includes only subhaloes resolved in Aq-A-1, it
is expected to increase by at most a factor of two if one extrapolates down to
the free-streaming length \cite[see][]{Springel2008}. The mass fraction in
substructures in this same region increases from $0.82\%$ for Aq-A-3 to
$3.3\%$ for Aq-A-1. Thus tidal streams contain almost $7$ times as much mass
as self-bound subhaloes at the resolution of Aq-A-1, and presumably would
contain even more at higher resolution. As Fig.~\ref{mass_radius} shows, the
substructure mass fraction varies as a function of radius. At the solar circle
about $0.05\%$ of the mass is in self-bound subhaloes and about $0.6\%$ in
tidal streams at the resolution of Aq-A-1. More than 99\% of the mass appears
smoothly distributed even at this extremely high resolution and when processed
with a state-of-the-art 6D structure finder.

\begin{figure}
\includegraphics[width=8.5cm,height=1.5cm]{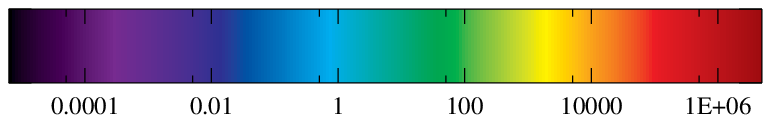}
\includegraphics[width=8.5cm,height=17.42cm]{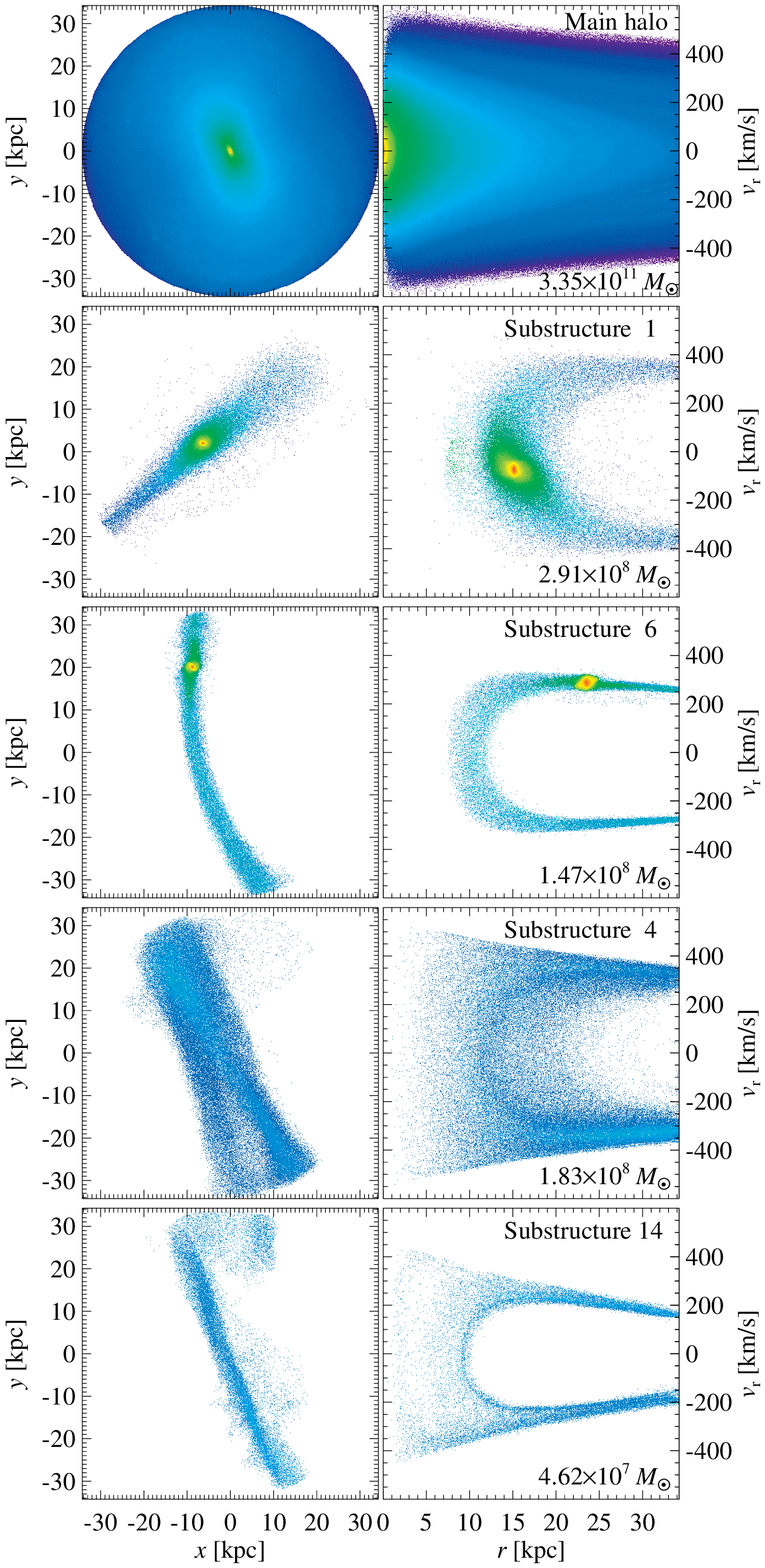}
\caption{Plots of the main halo and some selected substructures in $x$-$y$ and
  $r$-$v_r$ projections for Aq-A-1. We create a $500\times500$ image and colour
  each pixel according to the logarithm of the maximum 6D phase-space density 
  over the enclosed particles as estimated using EnBiD. The phase-space density
  is measured in units of $M_{\sun}~\mbox{kpc}^{-3}\mbox{km}^{-3}~\mbox{s}^3$.
  The mass of each substructure is given in units of $M_{\sun}$ at the bottom
  right of each row.}
\label{sub_rvr}
\end{figure}

To give a visual impression of the structures found by HSF, Fig.~\ref{sub_rvr}
shows some typical substructures in the inner halo of Aq-A-1. The top panel presents
the main halo with all HSF substructures (bound and unbound) removed. In the
second row we show the biggest bound subhalo and its attached tidal streams.
These extend over nearly 60~kpc. The mass of this biggest substructure is
$2.9\times10^8~{\rm M}_{\sun}$ of which $1.9\times10^8~{\rm M}_{\sun}$ is in
the self-bound subhalo. In the middle row we show an example of the same kind
of substructure, but here the tidal tails are more pronounced and the stream
passes within $8$~kpc of halo centre. The bottom two rows show two typical
stream-like structures which have no associated self-bound subhalo. Their
masses are similar to those of the biggest subhaloes.

\subsection{Density probability distribution function}
\label{sec:dens}

\begin{figure*}
\includegraphics[width=16cm,height=8cm]{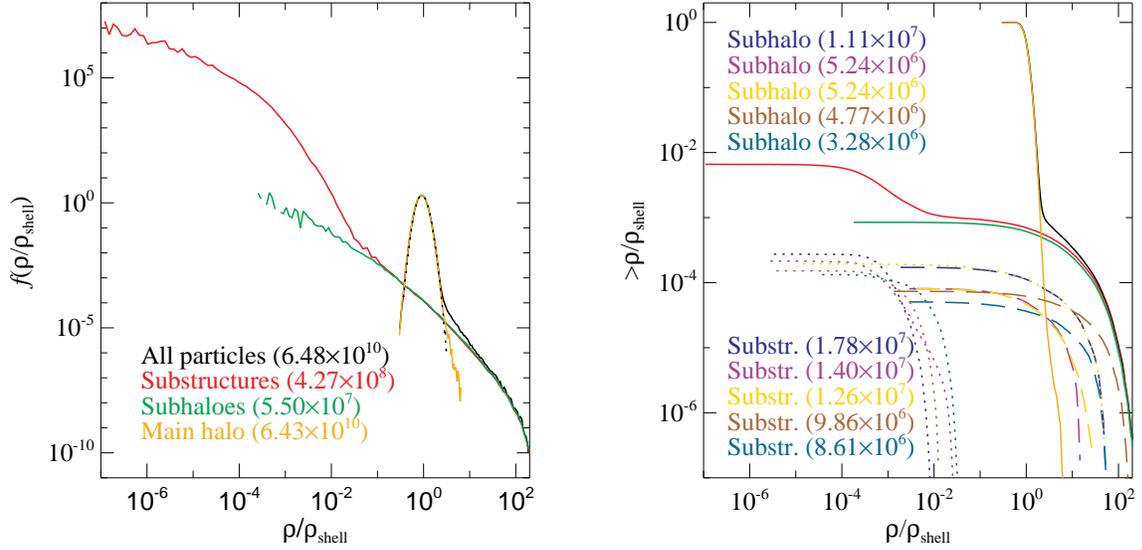}
\caption{{\it Left panel}: Volume-weighted density probability distributions
  for particles in the radial range $6-12$ kpc in Aq-A-1. For each particle an
  SPH-density is calculated using $32$ neighbours. The resulting density field
  is fitted to a smooth ellipsoidal power-law model to obtain $\rho_{\rm
    shell}$. SPH-densities are also calculated using the particles in each
  individual subhalo and substructure separately; we do not consider
  individual subcomponents containing fewer than $64$ particles. The results
  can be used to derive a density contrast $\rho/\rho_{\rm shell}$ at each
  particle's position, both in total mass and for the individual subcomponent
  to which the particle belongs. If $V$ denotes the total volume between 6 and
  12~kpc and $m_p$ is the particle mass, then $m_p/\rho V$ is the probability
  that a random point in this radial range overlaps the particle. By
  histogramming this quantity for all particles we obtain the probability that
  a random point has density contrast $\rho/\rho_{\rm shell}$ (the black solid
  line). By instead histogramming this quantity for the particles in a single
  component using the $\rho$ values calculated for individual subcomponents we
  obtain the mean number of subcomponents at a random position with local
  density contrast $\rho/\rho_{\rm shell}$ (yellow solid curve for the main
  subhalo, green for the set of self-bound subhaloes, red for the set of all
  substructures). Labels give the total mass in each component (in units of
  $M_{\sun}$). The black dashed line indicates the density contrast
  distribution produced by our density estimator for a Poisson realisation of
  a uniform density field. {\it Right panel}: Fraction of the mass in the
  radial range 6 to 12~kpc with density contrast above $\rho/\rho_{\rm shell}$
  for all particles, for the main subhalo, for the self-bound subhalo
  population and for the substructure population. For the latter two, the
  density contrast is that of the individual object containing the
  particle. Colours are as in the left panel. This panel also shows cumulative
  mass fraction plots for the five most massive self-bound subhaloes
  (long-dashed lines) and for the five most massive substructures (dotted
  lines). The masses of the individual objects are given in parentheses (in
  units of $M_{\sun}$).}
\label{volume}
\end{figure*}

In this section we try to understand how the different components contribute
to the dark matter density near the Sun. We ask how likely is it for the Solar
System to lie within a subhalo or tidal stream of given local density. This is
accomplished by computing probability density distributions for the space
density at random points within a thick spherical shell centred at the Solar
radius, as in \cite{Vogelsberger2009}, but separating the dark matter into the
different phase-space components identified by HSF. The result of this
procedure is presented in Fig.~\ref{volume}.

The left panel of Fig.~\ref{volume} shows the probability distribution function
of DM density for various structures in the inner halo. To make this plot we
first estimated the density at the position of every particle with radius
between 6 and 12~kpc using a standard SPH scheme based on 32 neighbours. As
described in \cite{Vogelsberger2009}, we then fitted a smooth model to these
values assuming the density to be stratified on similar, concentric ellipsoids and
to be a power law of radius. This defines a model density $\rho_{\rm shell}$
at the position of each particle which can be compared with the directly
estimated local density. This step is crucial to account for the large density
gradients in the inner halo so that we can focus on small-scale variations due
to substructure. We then repeat the SPH density estimates for the subsets of
particles in this radial range corresponding to each individual subcomponent:
the main subhalo and each individual self-bound subhalo and substructure. In
the following we plot all density distributions as functions of
$\rho/\rho_{\rm shell}$ and we construct volume-weighted probability
distributions by histogramming the particles with individual weights
$m_p/{\rho}V$, where $m_p$ is the particle mass and $V$ the total volume
between 6 and 12~kpc. For the total mass distribution and the main subhalo the
resulting distributions give the probability that an observer at a random
point in this radial range will see local density contrast $\rho/\rho_{\rm
  shell}$. For the self-bound subhalo and substructure components, the
distributions show the mean number of self-bound subhaloes or substructures
with local density contrast $\rho/\rho_{\rm shell}$ at a random point.

If we consider first the probability distribution of density contrast for the
total mass (i.e. the sum of all the components) we see that there is a
lognormal distribution centred on $\rho/\rho_{\rm shell}=1$ with an additional
low amplitude, power-law tail to high densities. This result was already given
in \cite{Vogelsberger2009}. As they showed, the lognormal part of the
distribution reflects discreteness noise in our density estimator. We
demonstrate this again here by plotting as a dashed black line the
distribution of analogous density estimates for points sampled from a uniform
Poisson distribution. This line cannot be distinguished from that
corresponding to the main subhalo in Fig.~\ref{volume}, demonstrating that
this component follows our ellipsoidal model very closely. The power law tail
is due to self-bound subhaloes, as is evident from its close agreement with
the distribution calculated directly from this component. This general
behaviour was pointed out not only in the analysis of this same simulation by
\cite{Vogelsberger2009}, but also in the analytic model of
\cite{Kamionkowski2008} and in the analysis of a different high resolution
simulation by \cite{Kamionkowski2010}.

If we now consider the density contrast distributions of the individual
components, we see that self-bound subhaloes are detectable not only in the
high-density tail but also down to contrasts as small as $10^{-4}$. This
reflects the excellent resolution of the Aq-A-1 simulation and, more
importantly, the fact that our 6-D structure finder can identify subhalo
material even at very low density contrast because of its small internal
velocity dispersion. HSF identifies general phase-space substructure
(e.g. tidal streams) down to even lower contrasts, of order $10^{-7}$. The
additional substructure mass which is not part of self-bound subhaloes is
almost entirely in this low-contrast regime. It is interesting that the
``probability'' a random point lies in such a low-density tidal stream reaches
values {\it much} larger than one, meaning that HSF has identified multiple
structures at each point in 3-space. Note, however, that because low-density
tidal streams have an effective spatial dimensionality less than 3, their
$\rho/\rho_{\rm shell}$ values are biased low (and the mean stream number
correspondingly high) by the spherical kernel of the SPH density estimator. In
the radial range between 6 and 12~kpc selected for this analysis about
$0.09\%$ of the mass is in the form of self-bound subhaloes and about $0.7\%$
in substructure. Thus low-density tidal streams account for almost 90\% of the
substructure detected by HSF.

The right panel of Fig.~\ref{volume} shows, for the various components, a
cumulative plot of the mass at local density contrast exceeding
$\rho/\rho_{\rm shell}$, expressed as a fraction of the total mass between 6
and 12~kpc. Here we see explicitly that the substructure component contains
almost ten times as much mass as the bound subhalo component and that this
excess lies almost exclusively at contrasts below 0.1. This panel also gives
similar cumulative data for the five individually most massive self-bound
subhaloes and for the five individually most massive substructures. Almost a
third of the mass in self-bound subhaloes is contained in these five objects,
almost all of it at density contrasts exceeding unity. However, only the most
massive subhalo corresponds to one of the five most massive substructures,
accounting for most of its mass. The other four massive substructures are
unbound tidal streams with no associated subhalo. The maximum density contrast
of these unbound streams is $\sim 10^{-2}$ (again this is probably biased low).
The five most massive substructures together account for only about 10\% of 
the total substructure mass.

\subsection{Velocity distributions}
\label{sec:vel}

\begin{figure}
\includegraphics[width=8.5cm,height=1.5cm]{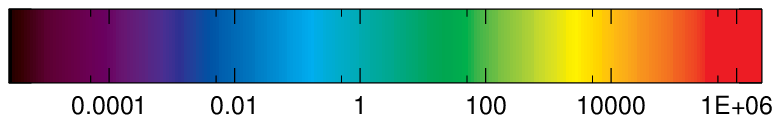}
\includegraphics[width=8.5cm,height=18.0cm]{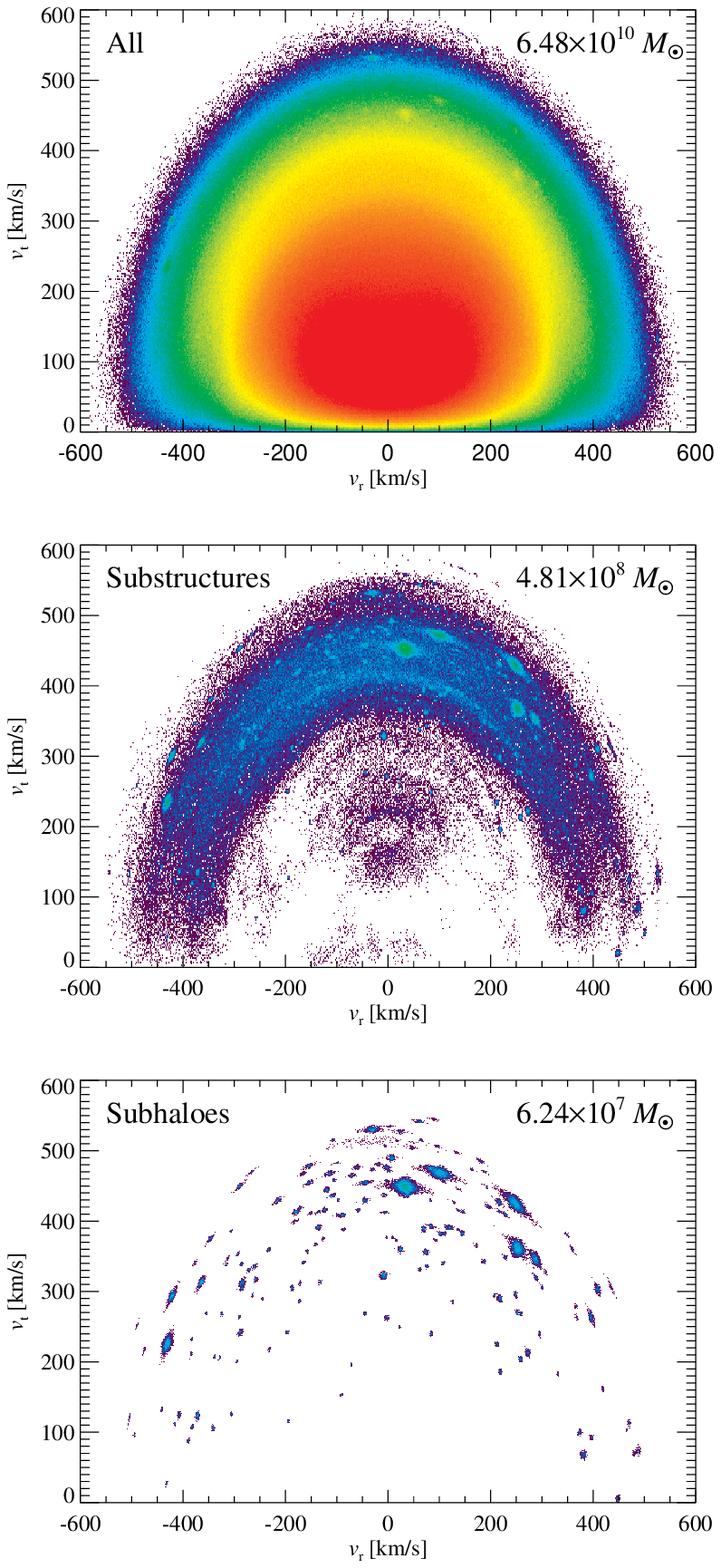}
\caption{Radial velocity $v_r$ -- tangential velocity $v_t$ phase-space plots
  for the $6-12$~kpc region of the Aq-A-1 halo. For each plot a
  two-dimensional histogram was calculated and colours were set to reflect the
  mass in each $2$~km~s$^{-1}$ x $2$~km~s$^{-1}$ pixel as shown by the colour bar,
  labelled in units of $M_{\sun}$. The top, middle and bottom panels show the
  distributions for all particles, for substructures and for subhaloes
  respectively.  The low velocity region is dominated by the main halo but
  some substructures/subhaloes are still visible in the the high $v_r$ and
  $v_t$ velocity regions. HSF only detects significant substructure in these
  regions of phase-space. The total component mass in $M_{\sun}$ is indicated
  in the top right-hand corner of each panel.}
\label{vrvt}
\end{figure}

Not only the mass density, but also the velocity distribution of DM particles
in the vicinity of the Earth is relevant for direct detection
experiments. \cite{Vogelsberger2009} showed that although this velocity
distribution is quite well approximated by a smooth trivariate Gaussian,
potentially measurable features are imprinted on the corresponding energy
distribution by the detailed formation history of the Milky Way's halo. Here
we concentrate on the velocities of the different phase-space components in
the inner halo. In Fig.~\ref{vrvt} we show $v_r$-$v_t$ projections of the
distribution of all particles in the radial range 6 to 12~kpc (top), of those
in substructures (middle), and of those in self-bound subhaloes (bottom).
These plots are two-dimensional histograms, with colour encoding the mass in
the corresponding bin as indicated by the colour bar (in units of solar masses
per $2$~km/s x $2$~km/s pixel). The total mass contributing to each panel is
given in its top right-hand corner. The main halo and so the bulk of the
particles lie primarily at velocities below 200~km~s$^{-1}$, whereas subhaloes
and tidal streams are found almost exclusively at higher velocities. As a
result, the most massive subhaloes are still (just) visible in the top panel
despite the fact that they contribute less than a tenth of a percent of the
mass. These structures contribute to the high-energy tail of the recoil
spectrum in direct DM detection experiments, and so may be visible in high
resolution experiments, particularly those with directional sensitivity which
can detect the common motion of the substructure particles.

\section{Conclusions}
\label{sec:con}

We study the population of subhaloes and tidal streams in six Milky Way-like
DM haloes taken from the Aquarius Project. These structures are identified
using the Hierarchical Structure Finder \citep[HSF][]{Maciejewski2009a}, a
state-of-the-art structure finder which operates in 6-D phase-space.

We find that that the differential mass function of self-bound subhaloes can
be well described by a power-law with slope close to $-1.9$. This agrees with
results from an independent analysis using the 3-D structure finder SUBFIND
\citep{Springel2008}. Typically HSF attaches slightly more particles to
subhaloes than SUBFIND, and also finds slightly more subhaloes above the
simulation resolution limit (see Table~\ref{table2}). This agrees with
previous results described in \cite{Maciejewski2009a}. About 14\% of the mass
within $r_{50}$ is in self-bound subhaloes, with significant scatter among the
six haloes Aq-A to Aq-F. HSF subhalo masses are $\sim 10\%$ larger than those
found by SUBFIND, although the increase can be larger near halo centre. In
most haloes the total subhalo mass is dominated by the largest objects. The
radial distributions of HSF and SUBFIND subhaloes are almost identical,
although HSF can identify subhaloes closer to halo centre due to their
enhanced density contrast in phase-space.

The differential mass function for substructures (i.e. both subhaloes and
tidal streams) is also well described with a power-law, but in this case the
slope is close to $-2$. This is independent of simulation resolution and holds
approximately for all six haloes, thus appearing robust. For most of the level
2 haloes around $35\%$ of the mass within $r_{50}$ is assigned to
substructures with mass above $\sim3\times10^5 M_{\sun}$. The radial
distribution of these objects can be approximated by equation (\ref{eq:fsub})
introduced in \cite{Springel2008} but with a higher normalisation than applies
for self-bound subhaloes. In the inner halo almost $10$ times as much mass is
detected in unbound tidal streams as in self-bound subhaloes. This reflects
the efficiency with which tidal forces destroy bound subhaloes in the inner
halo.

In our highest resolution halo, Aq-A-1, HSF assigns about $0.5\%$ of the mass
within $35$~kpc to self-bound subhaloes, and about $3.3\%$ to substructures
(subhaloes and tidal streams), with masses higher then $3\times10^4M_{\sun}$.
In this region the largest phase-space substructures are either self-bound
subhaloes with massive tidal tails stretching across the entire inner $35$~kpc
region or equally massive tidal streams with no attached subhalo at
$r<35$~kpc. The largest individual substructures in the inner region have
masses up to $3\times10^8M_{\sun}$.

\cite{Vogelsberger2009} showed that the density field in the radial range from
6 to 12~kpc is very well represented by a simple smooth model where density is
stratified on similar concentric ellipsoids and falls as a power law of
radius. Fluctuations around this model are small except for a low-amplitude
power-law tail to high density contrast which corresponds to self-bound
subhaloes. Our HSF analysis is consistent with these results and allows us, in
addition, to study the contrast of individual substructures with respect to
the smooth background, the main subhalo, in which they are embedded. HSF is
able to identify not only the high-contrast cores of individual self-bound
subhaloes, but also their outskirts where the density contrast drops to values
as low as $\sim 10^{-4}$. The maximum contrast of unbound tidal streams is
$\sim 10^{-2}$. Since these streams contain $0.6\%$ of the total mass between
6 and 12~kpc, of order one massive tidal stream is predicted to pass through
every point, contributing a few tenths of a percent of the local DM
density. In contrast, only $0.09\%$ of the mass in this region is contributed
by self-bound subhaloes, and the chance that the Earth lies in the
high-density contrast region of such a subhalo is below $10^{-4}$. Both
subhaloes and tidal streams populate the high-energy tail of the velocity
distribution preferentially, and would show up in direct DM detection
experiments as a small but significant part of the signal in events with
almost identical (vector) momenta.

\section{Acknowledgements}

The Aquarius Project was carried out as part of the activities of the
Virgo Consortium. We thank the consortium members responsible for
setting up and executing the simulations, as well as the consortium as 
whole for making them available to us.

\label{lastpage}

\end{document}